\documentclass[10pt,letterpaper]{article}
\usepackage[top=0.85in,left=2.75in,footskip=0.75in,marginparwidth=2in]{geometry}

\usepackage[utf8]{inputenc}

\usepackage{cite}
\usepackage{amsthm,amsmath}
\usepackage{amssymb}
\usepackage{nameref,hyperref}
\usepackage{hyperref}
\usepackage{float}
\usepackage{graphicx}
\graphicspath{{figures/}} 

\usepackage[right]{lineno}

\usepackage{microtype}
\DisableLigatures[f]{encoding = *, family = * }

\raggedright
\usepackage{changepage}

\usepackage[aboveskip=1pt,labelfont=bf,labelsep=period,singlelinecheck=off]{caption}

\makeatletter
\renewcommand{\@biblabel}[1]{\quad#1.}
\makeatother

\usepackage{lastpage,fancyhdr,graphicx}
\usepackage{epstopdf}
\fancyheadoffset[L]{2.25in}
\fancyfootoffset[L]{2.25in}

\usepackage{color}

\definecolor{Gray}{gray}{.25}

\usepackage{graphicx}

\usepackage{sidecap}

\usepackage{wrapfig}
\usepackage[pscoord]{eso-pic}
\usepackage[fulladjust]{marginnote}
\reversemarginpar

\begin{document}
\vspace*{0.35in}

\begin{flushleft}
{\Large
\textbf\newline{Effect of Neuromodulation of Short-Term Plasticity on Information Processing in Hippocampal Interneuron Synapses}
}
\newline
\\
Elham Bayat Mokhtari\textsuperscript{1},
J. Josh Lawrence\textsuperscript{2},
Emily F Stone\textsuperscript{1,*}
\\
\bigskip
\it{1} Department of Mathematical Sciences, The University of Montana, 32 Campus Dr, 59812, Missoula, USA.
\\
\it{2} Department of Pharmacology and Neuroscience, Texas Tech University Health Sciences Center, 3601 4th Street, 79430, Lubbock, USA.
\\
\bigskip
*stone@mso.umt.edu

\end{flushleft}

\section*{Abstract}
Neurons in a micro-circuit connected by chemical synapses can have their connectivity affected by the prior activity of the cells. The number of synapses available for releasing neurotransmitter can be decreased by repetitive activation through depletion of readily releasable neurotransmitter (NT), or increased through facilitation, where the probability of release of NT is increased by prior activation. These competing effects can create a complicated and subtle range of time dependent connectivity. Here we investigate the probabilistic properties of facilitation and depression (FD) for a presynaptic neuron that is receiving a Poisson spike train of input. We use a model of FD that is parameterized with experimental data from a hippocampal basket cell and pyramidal cell connection, for fixed frequency input spikes at frequencies in the range of theta and gamma oscillations. Hence our results will apply to micro-circuits in the hippocampus that are responsible for the interaction of theta and gamma rhythms associated with learning and memory. A control situation is compared with one in which a pharmaceutical neuromodulator (muscarine) is employed. We apply standard information theoretic measures such as entropy and mutual information, and find a closed form approximate expression for the probability distribution of release probability. We also use techniques that measure the dependence of the response on the exact history of stimulation the synapse has received, which uncovers some unexpected differences between control and muscarine-added cases.\\


\section*{Introduction}
Neuronal activity can have profound effects on functional connectivity at the level of synapses.  Through repetitive activation, the strength, or efficacy, of synaptic release of neurotransmitter (NT) can be decreased, through depletion, or increased, through facilitation.  Both of these competing processes involve intracellular calcium and may occur within a single synapse.  Different time scales of facilitation and depression enable temporally complex functional connectivity.  Here we investigate the probabilistic properties of facilitation and depression (FD) for a presynaptic neuron that is receiving repetitive in-vivo like Poisson spike train of input, e.g. the inter-spike interval follows an exponential distribution.  We use a model of FD that was parameterized with experimental data from dual whole-cell recordings from a presynaptic parvalbumin-positive (PV) basket cell (BC) connected to a postsynaptic CA1 (Cornu Ammonis 1 subregion) pyramidal cell, for family of fixed frequency input spikes into the presynaptic PV BC \cite{Lawrence2408}. Our results will thus apply to micro-circuits in the hippocampus that participate in the generation of theta and gamma rhythms associated with learning and memory.

The role of synaptic plasticity and computation has been analyzed and reported on in numerous papers over the past 30 years. A review of feed-forward synaptic mechanisms and their implications can be found in \cite{Abbott2004}. In this paper Abbott and Regher state ``The potential computational power of synapses is large because their basic signal transmission properties can be affected by the history of presynaptic and postsynaptic firing in so many different ways." They also outline the basic function of a synapse as a signal filter as follows: Synapses with an initial low probability of release act as high pass filters through facilitation, while synapses with an initially high probability of release exhibit depression and subsequently serve as low pass filters. Intermediate cases in which the synapse can act as a band-pass filter, exist. Furthermore, short-term plasticity can influence the availability of postsynaptic receptors to bind neurotransmitter.  For example, reducing neurotransmitter release probability can reduce postsynaptic receptor desensitization, effectively increasing the efficacy of synaptic transmission during high frequency stimulation.

Other functional roles of short-term synaptic plasticity can be found in \cite{ANWAR201771}.   They discuss signal processing capabilities in the auditory system, visual processing in the retina, olfactory processing and even electrosensory processing in weakly electric fish. In the hippocampus facilitation of excitatory synapses combined with depression of inhibitory synapses can amplify high frequency inputs, which selects for the high frequency output seen in place cells.  Inhibitory interneuron synapses in the hippocampus, such as the kind studied here, show a dynamic range of reaction when recruited via different pathways.  In a related synopsis \cite{Vogels2013}, the plasticity of inhibitory connections is studied in the context of spike timing dependent plasticity and network function.  Taking network function with plasticity a step further, \cite{Destexhe2004} analyzes simple neural networks in cortex with Hebbian-type learning rules affected by plasticity, modeling memory storage in networks with inhibitory synapses.  Results are largely general and any conclusions drawn are not immediately applicable to actual brain function.  Furthermore, the distinction between long and short term plasticity in this analysis is not made clear.

Synaptic depression as a regulator of synaptic transmission is discussed thoroughly in \cite{Abbott221}. Here it is supposed that synaptic depression can keep synaptic efficacy constant relative to changes in probability of release. Also, depressing synapses can serve as a gain control in the face of rapidly firing presynaptic cells.  It is also establishes that synaptic depression favors temporal encoding of information because the steady state is reached quickly and maintains no information about the absolute firing rate.  However, these synapses are sensitive to a sudden change in firing rate, detecting rate changes in low and high frequency inputs with similar sensitivity.

In cortex \cite{Fauth2014} draws the distinction between what they call synaptic vs. structural plasticity, focusing on structural plasticity, which directly effects the synaptic weights in a neural network model.  They find that certain characteristics arise directly from the interaction of structural plasticity and synaptic plasticity rules.  These characteristics in turn create a variety of stable synaptic weight distributions which could support information storage mechanisms. Neuromodulation can further alter the time dependent characteristics of a synapse.  Acting on the presynaptic side, many neuromodulators reduce the probability of release, protecting the synapse from depletion and therefore extending the duration or frequency sensitivity of the synapse overall.

The calculation of various kinds of measures of information transfer at the synapse level has been explored in many papers.  For instance, in \cite{CARTLING2002275} information transmission is studied through a master equation based stochastic model of presynaptic release of vesicles (which depends on intracellular calcium concentration), combined with a low dimensional model of membrane charging at the post-synaptic side.  The model itself has an advantage over models of average quantities, in that it captures fluctuations of the dynamic variables.  In \cite{Fuhrmann140} they consider synaptic transmission in neocortex, and compute the amount of information conveyed by a single response to a specific sequence of spike stimulation, as it is effected by short term synaptic plasticity.  They determine that for any given dynamic synapse there is an optimal frequency of input stimulation for which the the information transfer is maximal.  A mathematical model of the calyx of Held was used in \cite{Yang2014} to study synaptic depression due to repeated stimulation seen in vitro.  They quantify the information contained in the postsynaptic current amplitude about preceding interspike intervals using a mutual information calculation.  Both \cite{Fuhrmann140} and \cite{Yang2014} have directly inspired the work we present in this paper.

In \cite{Stone2014AKM} we parameterize a simple model of presynaptic plasticity from work by Lee and colleagues \cite{Lee2008} with experimental data from cholinergic neuromodulation of GABAergic transmission in the hippocampus. The model is based upon calcium dependent enhancement of probability of release and recovery of signalling resources. (For a review of these mechanisms see \cite{Khanin2006}). It is one of a long sequence of models developed from 1998 to the present, with notable contributions by Markram, \cite{Markram1998}, and Dittman, Kreitzer and Regehr \cite{Dittman1374}. The latter is a good exposition of the model as it pertains to various types of short term plasticity seen in the central nervous system, and the underlying dependence of the plasticity is based on physiologically relevant dynamics of calcium influx and decay within the presynaptic terminal.  In our work, we use the Lee model to create a two dimensional discrete dynamical system in variables for calcium concentration in the presynaptic area and the fraction of sites that are ready to release neurotransmitter into the synaptic cleft. The map is parameterized with experimental results from paired whole-cell recordings at CA1 PV basket cell-pyramidal cell synapses. The PV basket cell was presented with current pulses at fixed frequencies, resulting in trains of presynaptic action potentials evoking GABA transmission onto the postsynaptic pyramidal cell.  Synaptic transmission is manifested as trains of GABAA receptor-mediated inhibitory postsynaptic currents (IPSCs).  Experiments were run in control and with muscarine added, which acts upon presynaptic muscarinic acetylcholine receptors (mAChR) to cause a reduction in the observed IPSCs. Various parameterizations and hidden parameter dependencies were investigated using Monte Carlo Markov Chain (MCMC) parameter estimation techniques.  This analysis reveals that frequency dependence of cholinergic modulation requires both calcium-dependent recovery from depression and mAChR-induced inhibition of presynaptic calcium channels. A reduction in calcium entry into the presynaptic terminal in the kinetic model accounted for the frequency-dependent effects of the mAChR activation.

We now use our model to investigate the information processing properties of this synapse, in control and neuromodulation conditions. We possess these two parameterizations; one from experiments in control conditions, the other from synapses undergoing activation of mAChR receptors by muscarine.  Hence we can analyze the effect of this cholinergic modulation on information processing at the synapse level.  Because network oscillations associated with learning and memory feature PC basket cells firing in the gamma range (20 to 50 Hz) \cite{doi:10.1146/annurev-neuro-062111-150444}, we examine a range of frequencies from near zero to 100 Hz in what follows.

 As mentioned previously, this analysis is motivated and guided by the work of Markram and colleagues in \cite{Fuhrmann140}. In this paper it is analyzed how much information about previous interspike intervals is contained in the size of a single response of a dynamic synapses, i.e. a synapse which is affected by the history of presynaptic activity.  They derive expressions for the optimal frequency of the input Poisson spike train in terms of coding temporal information in depressing and facilitating neocortical synapses.  It was found that depressing synapses are optimal in this sense at low firing rates (0.5-5Hz) while facilitating synapses are optimal at higher rates (9-70 Hz).  This is not surprising, given that the average postsynaptic response is larger at low frequencies in depressing synapses, and larger at higher frequencies in facilitating synapses.  The more interesting result is the presence of a peak in the information transferred at a set frequency, an actual local maximum that is particular to each specific synapse.  We examine how cholinergic neuromodulation (the addition of muscarine) effects this result in what follows, and an alternative way to measure the information transferred from a sequence of preceding interspike intervals.

\section*{FD Model}
Many mathematical models have been developed to describe short-term plasticity over the past 20 years \cite{Tsodyks21011997, Markram1998, Lu, Dittman1374}. More flexible models include intracellular calcium dynamics \cite{doi:10.1146/annurev.physiol.64.092501.114547}, because probability of release is thought to be directly dependent upon calcium concentration in the presynaptic terminal. Recovery from synaptic depression is also thought to be accelerated by the presence of calcium, thereby unifying the underlying molecular mechanisms of facilitation and depression \cite{Dittman6147, Dittman1374, STEVENS1998415, vonGersdorff8137}.

In our model we take the probability of release ($P_{rel}$) to be the fraction of a pool of synapses that will release a vesicle upon the arrival of an action potential at the terminal. Following the work of Lee et al. \cite{Lee2008}, we postulate that $P_{rel}$ increases monotonically as function of calcium concentration in a sigmoidal fashion to
asymptote at some $P_{max}$. The kinetics of the synaptotagmin-1 receptors that binds the incoming calcium suggests a Hill equation with coefficient 4 for this function. The half-height concentration value, $K$, and $P_{max}$ are parameters determined from the data.

After releasing vesicles upon stimulation, some portion of the pool of synapses will not be able to release vesicles again if stimulated within some time interval, i.e. they are in a refractory state. This causes ``depression"; a monotonic decay of the amplitude of the response upon repeated stimulation. The rate of recovery from the refractory state is thought to depend on the calcium concentration in the presynaptic terminal \cite{Dittman6147, STEVENS1998415, Wang1998}.  Following Lee et al.,
 \cite{Lee2008}, we assume a simple monotonic dependence of rate of recovery on calcium concentration, a Hill equation with coefficient of 1, starting at some $k_{min}$, increasing to $k_{max}$ asymptotically as the concentration increases, with a half height of $K_{recov}$.  Muscarine, binding with muscarinic acetylcholine presynaptic receptors (mAChR), is thought to cause inhibition of presynaptic calcium channels, thereby decreasing the amount of calcium that floods the terminal when it receives an action potential. \cite{Gonzlez2014}.

\begin{figure}[H]
   \includegraphics[scale=1.2]{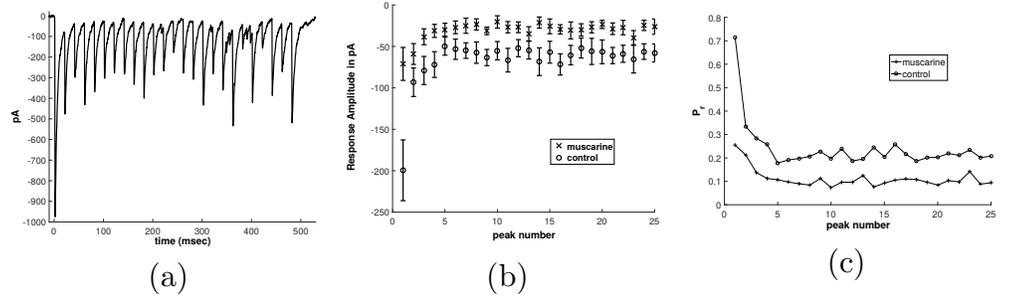}
  \caption{
      a) Response to pulse train stimulus from one cell in
control conditions, in pA, picoamps. b) Comparison of
response in control and muscarine conditions, averaged over 7 cells,
plotted with error bars at one standard deviation of the mean. The
baseline for each pulse has been subtracted from each peak, to
capture the {\it change} in the current upon stimulation. c) Absolute value of the response in control and muscarine conditions, baseline removed as
described in b), averaged over 7 cells and normalized by $N q$ ($N$ number of synapses,
$q$ quantal amplitude of single synapse) to
have units of ``probability of release", $Pr$.}\label{trains}
     \end{figure}

An example of this process is seen in the set of experiments illustrated in Figure \ref{trains}. Here whole-cell recordings were performed from synaptically connected pairs of neurons in mouse hippocampal slices from PV-GFP mice \cite{Lawrence2408}. The presynaptic neuron was a PV basket cell (BC) and the postsynaptic neuron was a CA1 pyramidal cell. Using short, 1-2 ms duration suprathreshold current steps to evoke action potentials in the PV BC from a resting potential of -60 mV and trains of 25 of action potentials are evoked at 5, 50, and 100 Hz from the presynaptic basket cell.  The result in the postsynaptic neuron is the activation of $GABA_A$-mediated inhibitory postsynaptic currents (IPSCs).  Upon repetitive stimulation, the amplitude of the synaptically evoked IPSC declines to a steady-state level.  These experiments were conducted with 5, 50 and 100 Hz stimulation pulse trains, in order to test frequency dependent short term plasticity effects. We note that oscillations in neural networks in the ``gamma" range (around 60-70 Hz) are associated with learning and memory, for a review see \cite{doi:10.1146/annurev-neuro-062111-150444}. Muscarine activates presynaptic metabotropic/muscarinic acetylcholine receptors (mAcHRs) which cause a reduction in the response overall, and subsequently the amount of
depression in the train.

The peak of the measured postsynaptic IPSC is presumed to be proportional to the total number of synapses that receive stimulation $N_{{tot}}$, which are also ready to release $(R_{rel})$, e.g. $N_{{tot}} R_{rel}$, multiplied by the probability of release $P_{rel}$. That is, peak IPSC $\sim N_{{tot}} R_{rel} P_{rel}$. $P_{rel}$ and
$R_{rel}$ are both fractions of the total, and thus range between 0 and 1. Without loss of generality, we consider peak IPSC proportional to $R_{rel} P_{rel}$.

The presynaptic calcium concentration itself, $[Ca]$, is assumed to follow first order decay kinetics to a base concentration, $[Ca]_{{base}}$. At this point we choose that $[Ca]_{{base}}=0$, since locally (near the synaptotagmin-1 receptors) the concentration of calcium will be quite low in the absence of an action potential.
The evolution equation for $[Ca]$ then is simply $\tau_{ca} \frac{d [Ca]}{dt}= - [Ca]$ where $\tau_{ca}$ is the calcium decay time constant, measured in
msec$^{-1}$. Upon pulse stimulation, presynaptic voltage-gated calcium channels open, and the concentration of calcium at the terminal increases rapidly by an amount $\delta$ (measured in $\mu m$):  $[Ca]\rightarrow [Ca]+\delta$ at the time of the pulse. Note that calcium build-up is possible over a train of pulses if the decay time is long enough relative to the inter-pulse interval.

The solution to the equation for calcium concentration can be simplified by defining a new time scale, $\tau={t}/{\tau_{{Ca}}}$. We also non-dimensionalize the calcium
concentration, rescaling it by the value of $\delta$ in the control case, $\delta_c$, and defining $C=\frac{[CA]}{\delta_c}$. After a stimulus occurring at a time $t=0$, which results in an increase in $C$ by an amount $\Delta=\frac{\delta}{\delta_c}$, the concentration of calcium is
$C(\tau)=C_0 e^{-\tau}+\Delta.$ In the control case this further simplifies to $C(\tau)=C_0 e^{-\tau}+ 1.$

As mentioned previously, the probability of release $P_{rel}$ and the rate of recovery, $k_{{recov}}$, depend monotonically on the instantaneous calcium concentration
via  Hill equations with coefficients of 4 and 1 respectively.  E.g. $ P_{rel}= P_{max} \frac{C^4}{C^4+K^4},$ and $ k_{{recov}} = k_{min} + \Delta k \frac{C}{C+K_{r}}.$
The variable $R_{rel}$ is governed by the ordinary differential equation $\frac{d R_{rel}}{dt}=k_{{recov}}(1-R_{rel}),$ which can be solved exactly for $R_{rel}(t)$.
$R_{rel}(t)=1-(1-R_0) (\frac{C_0 e^{-t} + K_r}{K_r +C_0})^{\Delta_k} e^{-k_{min} t}$ $P_{rel}$ is also a function of time as it follows the concentration of calcium after a stimulus.

We are interested in capturing the peak value of the IPSC, so we  construct a discrete dynamical system (or ``map") that
describes $P_{rel} R_{rel}$ upon repetitive stimulation. Given an inter-spike interval of $T$, the calcium concentration after a stimulus is $C(T) + \Delta$, and the peak IPSC is proportional to $P_{rel}(T) R_{rel}(T)$, which depend upon $C$. After the release, $R_{rel}$ is reduced by the fraction of synapses that fired, e.g. $R_{rel} \rightarrow R_{rel} - P_{rel} R_{rel}= R_{rel}(1-P_{rel})$. This value is used as the initial condition in the solution to the ODE for $R_{rel}(t)$.  A two dimensional map (in $C$ and $R$) from one peak value to the next is thus constructed. To simplify the formulas we let $P=P_{rel}$ and $R=R_{rel}$.  The map is
\begin{align*}
C_{n+1}&=C_n e^{-T}+\Delta\\
P_{n+1}&= P_{max} \frac{C_{n+1}^4}{C_{n+1}^4+K^4}\\
R_{n+1} &= 1-(1-(1-P_{n}) R_n)(\frac{C_n e^{-T}+K_r}{K_r+C_n})^{\Delta k} e^{-k_{min}T}
\end{align*}
The peak value upon the $n$th stimulus is $R_n P_n$, where $R_n$ is the value of the reserve pool before the release reduces it by the fraction $(1-P_n)$.

\subsection*{Parameter estimation}
The parameter values for the model are summarized in Table \ref{tble:1}. The rescaled data presented in a previous section were fitted to the map using the Matlab package lsqnonlin, and with MCMC techniques(\cite{haario2001,Haario2006}).  The value of $P_{max}$ was determined by variance-mean analysis, and is 0.85 for the control data and 0.27 for the muscarine data.  The common fitted parameter values for both data sets are shown in Table \ref{tble:2}.

The control data set was assigned $\Delta = 1$ which can be done without loss of generality if the concentration of calcium is scaled by that amount, and the muscarine data set has the fitted value of $\Delta = 0.17$. From this result it is clear that the size of the spike in calcium during a stimulation event must be much reduced to fit the data from the muscarine experiments. This is in accordance with the idea that mAChR activation reduces calcium ion influx at the terminal.

\subsection*{Discussion of the model}
It is common in the experimental literature to classify a synapse as being ``depressing" or ``facilitating", depending upon its response to a pulse train at some relevant frequency.  Simple models can be built that create each effect individually. The model here combines both mechanisms so that, depending upon the parameters, both facilitation and depression and a mixture of the two can be represented \cite{Lee2008}. Note that facilitation is built into this model through the calcium dependent $P$ value and rate of recovery.

This interplay of the presynaptic probability of release and the rate of the recovery creates a non-linear filter of an incoming stimulus train. Adding muscarine modifies the properties of this filter. To investigate this idea, we consider the distribution of values of $Pr$ created by exponentially distributed random ISIs for varying rates $\lambda$, or mean ISI, denoted $<T>=1/\lambda$. Doing so explores the filtering properties of the synapse when presented with a Poisson process spike train. To develop our intuition, we first present results from numerical studies to determine of the effect of varying frequency of the pulse train, or mean rate, on the muscarine and control cases. Then we create analytic expressions for the distribution of calcium and the $Pr$ values, which are corroborated by the numerical results. Finally the information processing properties of the synapse in the control and the muscarine cases at physiological frequencies are compared.

\section*{Numerical Study of the Distributions in $Pr$}
To create approximations to the distribution of $Pr$ values we computed $2^{14}$ samples from the stochastic map, after discarding a brief initial transient. The values, ranging between 0 and 1, were placed into evenly spaced bins.  The histograms, normalized to be frequency distributions, were computed for a range of mean frequencies (or rates) in the theta range, gamma range and higher (non-physiological, for comparison).  The parameters used in the following simulations are from fitting the model to the control and muscarine data set (see Table \ref{tble:2}).

\subsection*{Frequency distributions of $Pr$ in control and muscarine cases}
There is a general progression of the frequency distribution for $Pr$ as the rate of the incoming Poisson spike train is increased, in both cases. For very small rates (between almost 0 and 1 Hz) the distribution is peaked near $P_{max}$. The peak is skewed to the left necessarily, as it is restricted to the support $[0,1]$  and the exponential distribution of ISIs always contributes some very small values, which generate lower $Pr$ values. For very large rates (non-physical, 200 Hz and larger) the distribution is peaked near 0, reflecting the fact that the synapse does not have time to recover between spikes.  Again the peak is skewed, this time to the right, necessarily. In between the two extremes the distribution must spread out between 0 and 1, and it does so in a very particular way.

\begin{figure}[H]
   \includegraphics[scale=0.77]{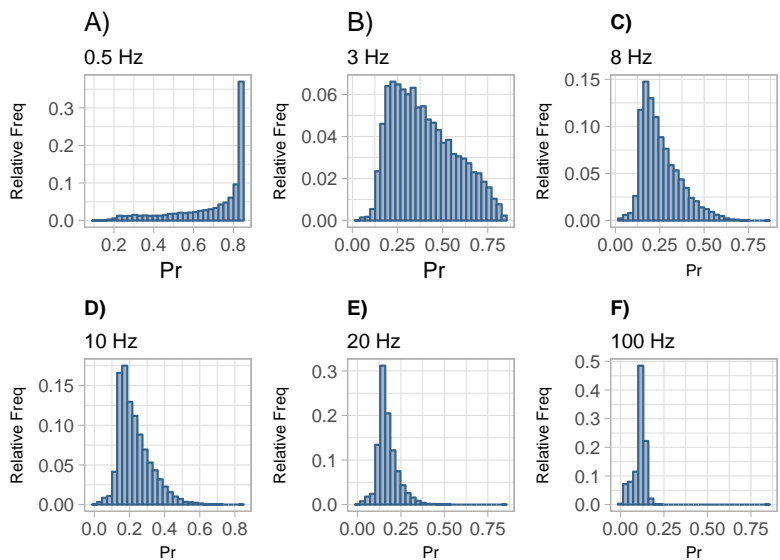}
  \caption{
      Frequency distribution of $Pr$ with control parameter set under stimulation at a) 0.5 b) 3 c) 8 d) 10 e) 20 and f) 100 Hz.  Horizontal axis is the $Pr$ value, vertical axis is the relative frequency of the $Pr$ value.}\label{FreqDist2}
      \end{figure}

As the rate is increased from 0.5 to 10 Hz the distribution spreads quickly out over the entire interval,  see Figure \ref{FreqDist2}. This might be expected, since the range of frequencies present in the exponential distribution will cause a wide range of $Pr$ responses at these lower frequencies. First the skewness to the left becomes more pronounced, then the left half of the distribution grows up to match the right, becoming pretty much flat at around 1.8 Hz.  It then begins to drop on the right side, becoming almost triangular around 3 Hz.  From there the peak on the left sharpens, while maintaining a shoulder for the very lowest values of $Pr$.  Note that this covers the range of rates generally thought to be theta frequency. Here the synapse could be said to be the most sensitive and allow for widely varying responses.

For rates larger than 10 Hz  the peak on the left sharpens as the rate is increased, but very slowly.  The shoulder on the left of the peak remains in place.  In contrast with the theta range, the gamma range is marked by a nearly steady response.  Having these two distinct ``tunings" of the synapse at physiological frequencies is quite interesting. For even larger frequencies the shoulder on the left grows up to meet the peak (data not shown).  The peak near 0.1 persists until the rate is greater than 300 Hz, after which it is subsumed into the peak near zero.  The synaptic dynamics thus creates a small probability response that is stable over a wide range of frequencies.

\begin{figure}[H]
   \includegraphics[scale=0.8]{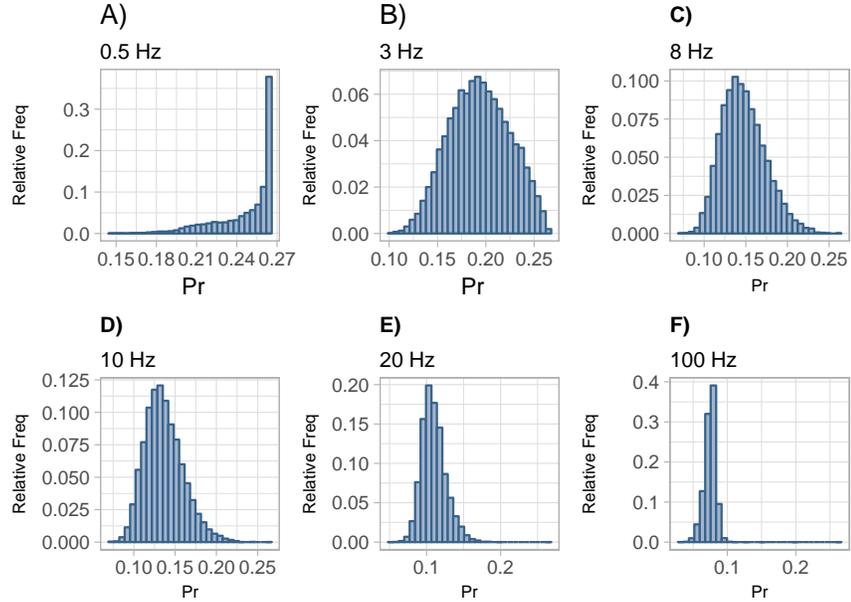}
  \caption{
     Frequency distribution of $Pr$ with muscarine parameter set under stimulation at a) 0.5 b) 3 c) 8 d) 10 e) 20, and f) 100  Hz.  Horizontal axis is the $Pr$ value, vertical axis is the relative frequency of the $Pr$ value.}\label{FreqDist2musc}
      \end{figure}

For the control case the influx of calcium ($\Delta$) was set to unity, because the variable in the map for calcium was rescaled by this amount.  In order to fit the muscarine data a much smaller value of ($\Delta$) was needed (0.17).  This is consistent with the hypothesis that muscarine shuts down the influx of calcium ions to the presynaptic cell.  This reduces the size of the response, but it was also shown to reduce the relative amount of depression seen with repeated spikes, at least at intermediate frequencies around gamma.  This could have important implications for the effect of neuromodulation at these frequencies.  How is this manifested in the $Pr$ distributions?  In Figure \ref{FreqDist2musc} are shown the $Pr$ frequency distributions for varying mean rate.  The range on the $x$-axis is set to $[0,0.3]$ because $P_{max} = 0.27$.  With this rescaling the distributions look much like those in the control case:  The distribution is skewed left with a peak near $P_{max}$ for low frequencies, and shifts to have a peak in the middle of the range for intermediate frequencies while spreading out.  The peak in the middle of the range gradually moves to the left as the frequency is increased further.  In the control case the peak is more triangular and skewed right than in the muscarine case at low frequencies, at gamma frequencies the control distribution has a shoulder on the left, while the muscarine case distribution is more symmetrical. Therefore the muscarine treated synapse focuses the response in a narrow range around a small $Pr$ for all but the lowest frequencies.  At high frequencies the peak of the muscarine distribution does not go to zero as fast as the control distribution does, a somewhat non-intuitive result, though we must point out that this range of frequencies is not physiological.

In both cases the dynamics of the map create a low pass filter, which is the hallmark of a depressing synapse.  The mid-range peak makes it something more complex than a linear low pass filter, as it has a typical response size (the peak or the mean value) for each frequency. However, it shows an interesting uniformity: the peak (or most likely) response is fairly insensitive to changes in frequency over the physiological range of 3-70 Hz.

\begin{figure}[H]
   \includegraphics[scale=1.2]{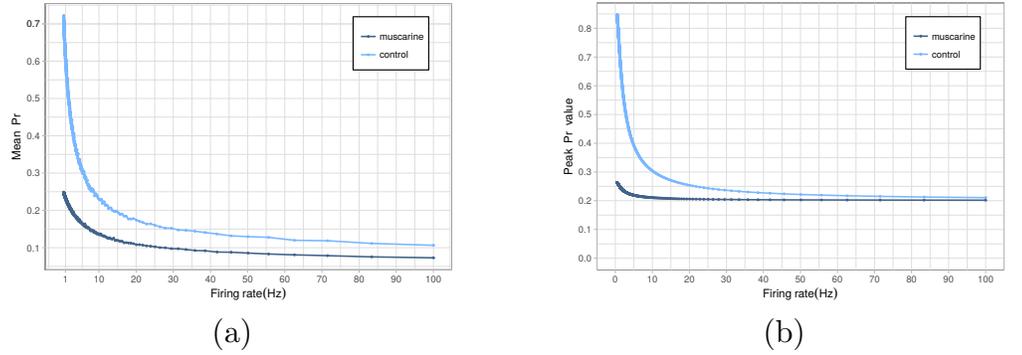}
  \caption{
     Mean (a) and peak (b) of the $Pr$ distribution vs. rate.
}\label{MeanPeakPvsRate}
      \end{figure}

We can then compare the mean and the peak of the distributions as the input frequency is varied.  See Figures \ref{MeanPeakPvsRate}. As expected, the mean and the peak of the muscarine distributions are much lower than that of the control, except in the case of the peak within the theta range.  This could mean that even under depression caused by pharmaceutical applications the synapse response remains consistent, a kind of built-in stability to theta frequencies.  We also note that the amount of depression relative to the initial response of the synapse is less for the muscarine case than the control: the mean ranges from over 0.8 to between 0.1 and 0.2 for the control case, a change of about 0.6-0.7 over all, and from 0.3 to near 0.1 in the muscarine case, a much smaller change of about 0.2.  This confirms the assertion in \cite{Lawrence2408}.

\subsection*{Entropy of $Pr$ distribution vs. mean rate}
The entropy of a distribution is a convenient scalar measure for comparing the overall structure in distributions as a parameter is varied.  We therefore compute the entropy of the $Pr$ distributions in the control and muscarine case for varying mean rate and compare it with the distributions themselves.  Finally, we draw some physiological conclusions.

The entropy of a distribution, $p(x)$, of a random variable, $X$, is defined to be:
\begin{eqnarray}
H\left(X\right) &= -\sum_{x\in\mathcal{X}}{p\left(X=x\right)\log_{2}p\left(X=x\right)}\label{margent}
\end{eqnarray}
It measures the number of states the variable can assume, along with the probability of each state. Note that for continuous random variables it is necessarily dependent upon the exact partition of the distribution into a histogram of values that can be summed.  In what follows we keep this partition constant across different cases, comparing the entropies of the resulting histograms. All other things being equal, the entropy of a distribution with more ``spread" is higher than one that is more focused on a single value. If a random variable is tightly constrained, its distribution will have a lower entropy, i.e. it is much more certain what state the variable will be in for any given sample.

\begin{figure}[H]
   \includegraphics[scale=1.2]{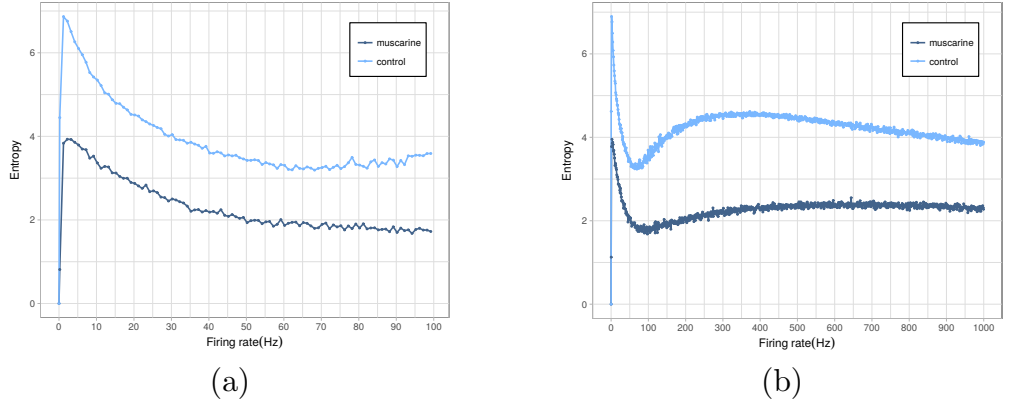}
  \caption{
    Entropy of the $Pr$ distribution vs. frequency for control and muscarine parameter sets.}\label{PrEntropy}
      \end{figure}

In Figure \ref{PrEntropy} we plot the entropy as a function of the mean rate of the driving exponential distribution for both control and muscarine parameter sets.  Figure \ref{PrEntropy} b) shows a range of frequencies from near zero to 1000 Hz to see the limit for large frequencies.  Figure \ref{PrEntropy} a) shows a zoom in on the physiological range of frequencies between zero and 100 Hz.

Studying the entropy vs. frequency from 0.1 to 100 Hz, we see a local maximum for low values of frequency (at approximately 4 Hz). In this frequency range the distribution spreads out between a peak at high $Pr$ values to a peak at lower $Pr$ values. At these frequencies there would be a large variability in the size of the response, which could be a useful characteristic in a communication channel.  However, if the criteria is a stable level of connectivity, lower entropy is desirable. For larger frequencies (not physiological) the entropy increases again to a local maximum near 260 Hz, after which it decays as the distribution narrows near almost zero $Pr$ values.  For the muscarine case the second local maximum is less pronounced and occurs near 388 Hz. The maximum in both cases is the result of the distribution shifting from one peaked near $Pr > 0$ to one peaked at zero skewed to the right. This occurs through the widening out of the peak, causing the increase in entropy.

Note that the size of response ($Pr$) is a measure of the ``strength" of the synaptic connection created by the ``pool" of synapses.  The advantage of having a peaked distribution with low entropy over a range of frequencies is that a stable connection is created, even when presented with a stochastic signal of the Poisson type.  Alternatively, when the distribution is switching from a peak at high $Pr$ values to low, as the mean rate is increased, the entropy is at a maximum and there is a greater range of coupling strengths.  In that case the exact value of the strength of the connection will depend on the past synaptic signaling history.

\section*{Stochastic Recurrence Relationships for Calcium and $Pr$}
The map for $C, P$ and $R$ driven by a Poisson spike train is a stochastic recurrence relation with noise added in an exponent.  We shall show that while it is not possible to create a closed form for the complete map, an approximation can be created that relies on the properties of the deterministic map for low spike train rates.  Only the map for the calcium concentration yields to direct analysis, but involves the introduction of a random variable for $\Delta$.  We provide an outline of the work required to create this, for completeness.

\subsection*{Calcium concentration}
Suppose an independent, identically distributed increase in the amount of calcium $\{\Delta_{n}\}_{n\geq1}$ occurs at times $\{t_n\}_{n\geq1}$, where the times are exponentially distributed with rate parameter $\lambda$.  We are interested in the distribution of the concentration of calcium accumulated in the presynaptic terminal following this supposition.

The concentration is governed by a random recurrence equation given by
\begin{eqnarray}
C_{n} = A_{n}C_{n-1}+\Delta_{n},\hspace{5mm}n\geq1,\label{Cn}
\end{eqnarray}
where $A_{n}= e^{-T_{n}/\tau_{ca}}$, and the waiting times $T_1=t_1$, $T_n = t_{n}-t_{n-1}$, $n\geq2$, are i.i.d., making $\{t_n\}$ a renewal process. Moreover, $\left(A_{n},\Delta_{n}\right)$ are assumed to be i.i.d. vectors with initial condition $C(0)=C_{0}$.  $C_{0}$ is a base concentration of calcium which is considered to be zero in the absence of a stimulus.  After some manipulation it can be shown (see Appendix A) that the concentration of calcium follows a Gamma distribution with a shape parameter $\lambda\tau_{Ca}+1$ and a scale parameter 1. Thus,
\begin{eqnarray*}
f_{C}\left(c\right) = \frac{1}{\Gamma(\lambda\tau_{ca}+1)}c^{\lambda\tau_{ca}}e^{-c};\hspace{5mm} \hspace{1mm}\lambda\tau_{ca}+1>0
\end{eqnarray*}
where $c$ is a realization of random variable $C$. The distribution of calcium concentration $C$ is centered around $\lambda\tau_{ca}+1$. The coefficient of variation of $C$ is $1/(\sqrt{\lambda\tau_{ca}+1})$, and hence, the shape of distribution depends on the value of $\lambda\tau_{ca}+1$. The details of the derivation can be found in the Appendix B.

\begin{figure}[H]
   \includegraphics[scale=0.65]{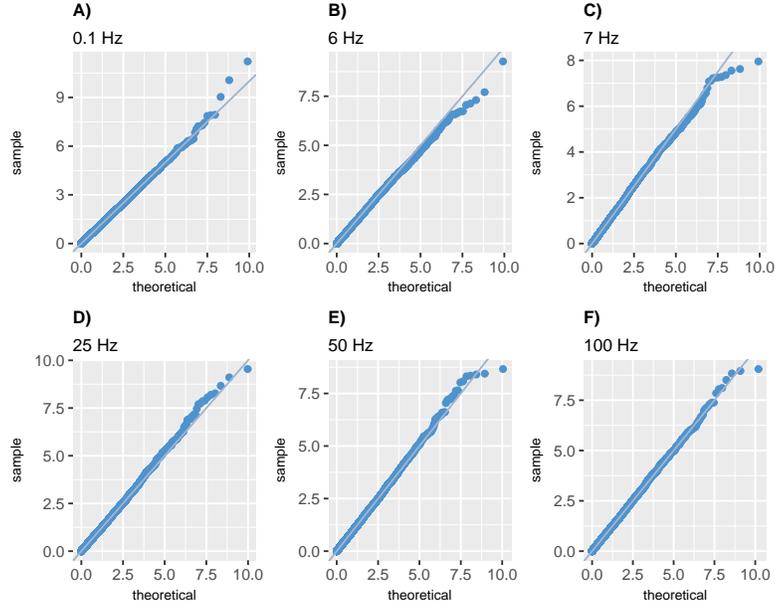}
  \caption{
    Fitted percentiles of gamma-distributed data to observed calcium concentration data from physical model for frequencies a) $0.1$ b) $6$ c) $7$ d) $25$ e) $50$ f) $100$  Hz. The line of identity is also plotted. The alignment of the line and quantile plot indicates the adequacy of fit of the gamma model for calcium concentration.}\label{gama}
      \end{figure}

We compare this result with the distribution obtained by numerical simulation of the recurrence relation for $C$ by creating quantile plots. Figure \ref{gama} displays quantile plots for the map with input frequencies  $0.1$, $6$, $7$, $25$, $50$, $100$  Hz, with the theoretical quantiles based upon the gamma distribution. This type of graphical display shows whether the simulated data can reasonably be described by a gamma distribution. Plots show adherence to a linear relationship between the observed and theoretical quantiles, confirming our analytic results.

\subsection*{Distribution of $Pr$ for small $\tau_{ca}$ and large $T$}
\begin{figure}[h!]
   \includegraphics[scale=0.65]{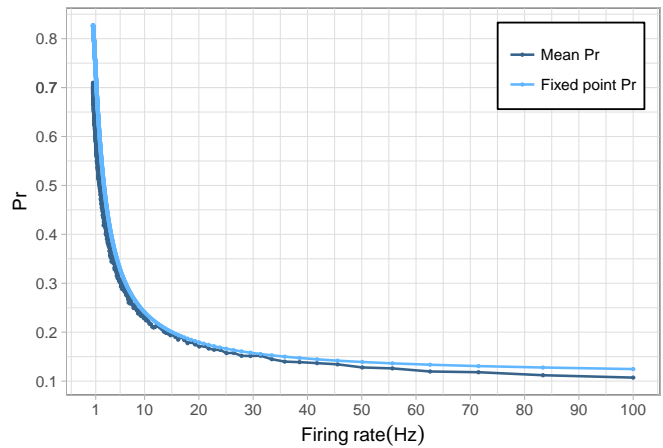}
  \caption{
     Fixed point values for the deterministic map vs. frequency, compared with the mean values for the computed distribution. }\label{FixedPoint}
      \end{figure}

We conclude from the previous subsection that the random variable describing the calcium concentration does have a closed form distribution, and indeed a well-known distribution.  However, this is not the case for the variable $R$ due to the complexity of the map, and so a closed form for the distribution of $Pr = P R$ is not possible.
However, we can understand it partially by considering the mechanisms involved, as done in the preceding section.  We can also use some information from the deterministic map.  The map has a single attracting fixed point, and the collapse to this fixed point from physiological initial conditions is very rapid \cite{Stone2014AKM}.  The value of the fixed  point depends on the frequency, with a smaller value for larger frequency in general. In Figure \ref{FixedPoint} we plot the expression for the fixed point of the deterministic map vs. rate, along with the mean of the distribution of $Pr$ for varying frequencies.  The values decrease with increasing frequency, as expected, and are remarkably close.

This motivates the idea that if the $Pr$ value was directly determined by the fixed point value for the ISI value preceding it, we would be able to convert the distribution of the ISIs into that of the $Pr$s by using composition rules for distributions of random variables. We examine this when the calcium decay time ($\tau_{ca}$) is notably smaller than the inter-spike interval ($T$).  With this approximation $C, P$ and $R$ have time in between pulses to decay to their steady state value before another pulse.  This means that the fixed point value for a rate given by $1/T$ where $T$ is the preceding interspike interval is more likely to give a good estimate of the actual value or $Pr = P R$.

It was shown in \cite{Stone2014AKM} that in this case $\bar{C}\rightarrow{\Delta}$ as $T$ increases and hence $\bar{P}\rightarrow{P_{max}}$.
Therefore, the fixed point $\bar{R}$ is then
\[\bar{R} = \frac{1-e^{-k_{min}T}}{1-(1-P_{max})e^{-k_{min}T}}.\]
From this we can compute the probability density function of $\bar{R}$, given an  Exponential distribution of the variable $T$.  Details of this calculation are provided in Appendix C. If $X = \bar{R}$ is a random variable, then an analytic expression for its probability density function (PDF) exists and is given by
\begin{eqnarray}
f\left(x|\lambda,c,k_{min}\right) = \frac{\lambda (1-c)}{k_{min}}(1-x)^{-(1-\lambda/k_{min})}(1-cx)^{-(1+\lambda/k_{min})},\label{Rbardist}
\end{eqnarray}
where $c=1-P_{max}$, $\lambda>0$ is the mean Poisson rate and $k_{min}>0$ is the baseline recovery rate. The distribution is supported on the interval $[0,1]$.

From the expression \ref{Rbardist}, the mean or expected value of random variable $X=\bar{R}$ is given by
\begin{align}
E\left(X\right) = (1-c)\lambda\left(\frac{1}{\lambda(1-c)}-\frac{_2F_1\left(1,\frac{k_{min}+\lambda}{k_{min}};2+\frac{\lambda}{k_{min}};c\right)}{k_{min}+\lambda}\right),
\end{align}
where $_2F_1\left(1,\frac{k_{min}+\lambda}{k_{min}};2+\frac{\lambda}{k_{min}};c\right)$ is the hypergeometric function.

\begin{figure}[H]
   \includegraphics[scale=0.55]{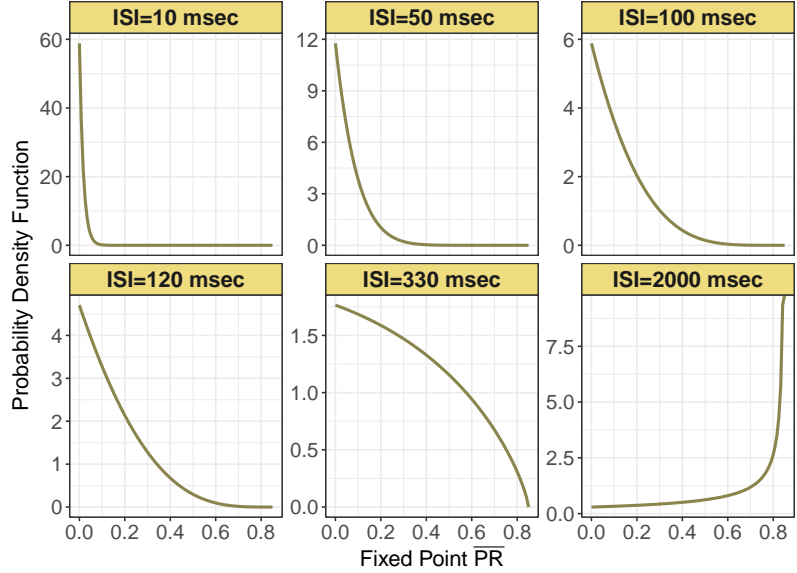}
  \caption{
      PDF of the stochastic fixed point $\overline{PR}$ for varying mean input ISI. Parameters $k_{min}=0.0013$ and $p_{max} = 0.85$, are from the control set.}\label{PRbardist}
      \end{figure}

Similarly, we can compute the analytical expression of the probability density function of fixed point $Y=\overline{PR}$.  We will refer to this in what follows as the {\it stochastic fixed point}. Hence, the probability density function for the stochastic fixed point is
 \begin{eqnarray}
f\left(y|\lambda,c,k_{min}\right) = \frac{\lambda P_{max}(1-c)}{k_{min}}(P_{max}-y)^{-(1-\lambda/k_{min})}(P_{max}-cy)^{-(1+\lambda/k_{min})}.\label{PRbardistrib}
\end{eqnarray}
This distribution is supported on the interval $[0,P_{max}]$.
Figure \ref{PRbardist} shows this expression for different mean input inter-spike interval, in milliseconds.

\begin{figure}[H]
   \includegraphics[scale=0.55]{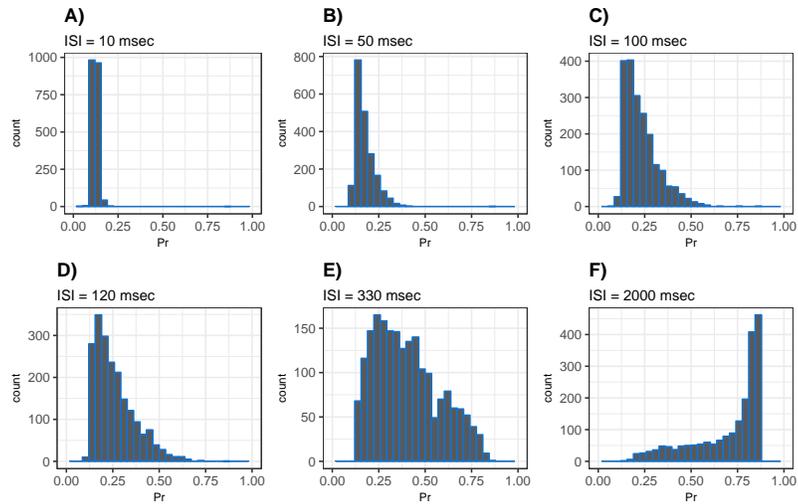}
  \caption{Distribution of $Pr$ for varying mean input ISI a)  $10$ b)  $50$ c) $100$ d) $120$ e) $330$ and f) $2000$ in milliseconds. }\label{IPSCLargeT}
             \end{figure}

In Figure \ref{IPSCLargeT} are histograms of $Pr$ values obtained from the map with very small $\tau_{ca}$, and other parameters from the control set, as in Figure \ref{PRbardist}.  The similarity between the two is evident.  Apparently this approximation captures not only the mean value of the numerical distribution, but also the shape of the distribution and how it changes with varying input spike train rate.

\begin{figure}[H]
   \includegraphics[scale=0.65]{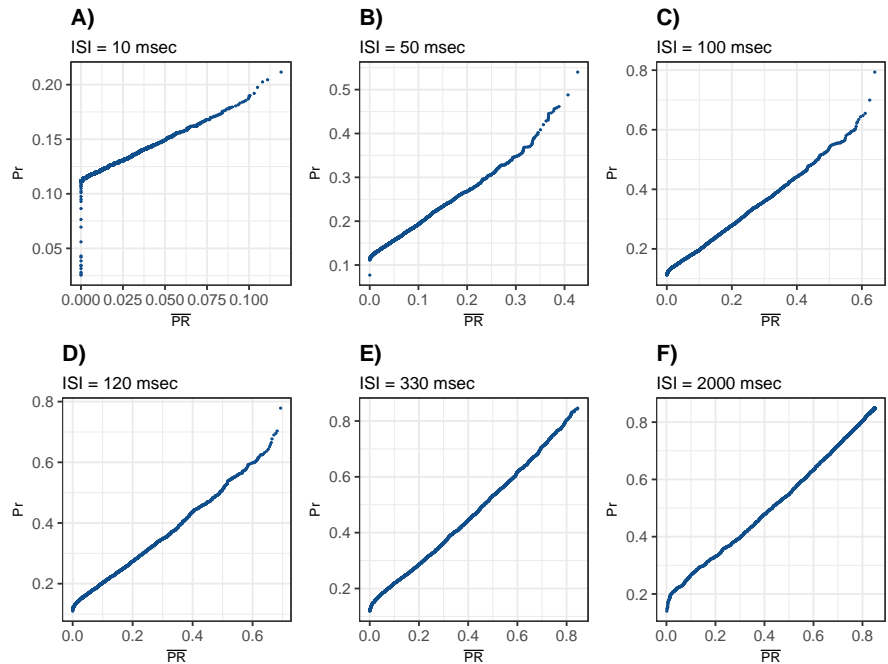}
  \caption{ Quantile plot of two data sets obtained from $Pr$ and $\overline{PR}$. }\label{qqplotfixedpointpr}
      \end{figure}

Figure \ref{qqplotfixedpointpr} compares these two distributions via quantile-quantile plots, for mean ISI of $10$, $50$, $100$, $120$, $330$, and $2000$  milliseconds. In every QQ-plot the quantiles of all $\overline{PR}$ are plotted against the quantiles of all $Pr$ values. If the values of the two different data sets have the same distribution, the points in the plot should form a straight line. From these plots it is clear that when the mean ISI is significantly larger than calcium decay time, the distribution of the stochastic fixed point is similar to that of $Pr$.  However, for smaller mean ISI ($10$ msec) the approximation becomes less exact, so the similarity between two distributions decreases.

\section*{The Stochastic Model of the Postsynaptic Response}
To compute mutual information between the stimulus and the response of a synapse we must capture the variability in the postsynaptic response.  A certain probability of release will generate a distribution of responses that has an analytical description given certain fundamental assumptions about stochastic processes.  We provide a short overview of these here.

\subsection*{Model description}
Noise is a fundamental constraint to information transmission, and such variability is inherent in individual neurons in the nervous system. To account for the variability across identical stimulation trials, we use stochastic model of the synapse.  Here, we follow the Katz model of neurotransmitter release \cite{STEVENS199355}.

Upon the arrival of an action potential at a presynaptic terminal, calcium influx through calcium channels causes some of the synaptic vesicles fuse to terminal bouton membrane at special release sites and diffuse their neurotransmitters across the synaptic cleft. Each site can release either one or zero vesicles and we assume $N$ release sites. Each release event is independent, and we can assume that release of $K$ vesicles from $N$ release sites follows \textit{Binomial} distribution with the two parameters ($N$, $Pr$) and is written
\[K\sim \mathcal{B}(N,P_r)\]
$Pr$ is the release probability for each release site following an action potential. Note that failure of release results in a zero amplitude response from the postsynaptic neuron, so it cannot be informative.
In addition, there is not only variability in the number of sites activated and hence the number of vesicles actually released, but also in the postsynaptic response to a single vesicle, due to inherent stochasticity in the postsynaptic receptors. Therefore, we assume that the size of the postsynaptic response ($\mathcal{S}_{resp}$) at the time of spike is not a constant, but instead follows \textit{Normal} distribution with mean $\mu$ and variance $\sigma^2$, with a two-sided truncation which is written
\[\mathcal{S}_{resp_{i}}\sim \mathcal{N}(\mu,\sigma)\]
Therefore, the amplitude of postsynaptic response following each action potential is obtained by combining the Binomial model of vesicle release with the Normal model of a single response. The summation of responses evoked by each release can be written
\[ \mathcal{S}= \left\{
 \begin{array}{l l}
   0 &     \text{if}\;\;K=0,\\
  \sum_{i=1}^{k}{\mathcal{S}_{resp}}_{i}&    \text{if}\;\; K>0.
\end{array} \right.\]
Note that for $K>0$, $\mathcal{S}\sim N(k\mu,\sqrt{k}\sigma)$.
The probability density of the postsynaptic response to release of single vesicle is thus
\begin{eqnarray*}
 f(\mathcal{S}={\scriptstyle\mathcal{S}}|\mu,\sigma^2)= \left\{
 \begin{array}{l l}
   \frac{1}{\sqrt{2\pi\sigma^2}\mathcal{N}_{cts}}e^{-\frac{({\scriptstyle\mathcal{S}}-\mu)^2}{2\sigma^2}} &     \text{if}\;\; 0<{\scriptstyle\mathcal{S}}<2\mu,\\
 0&    \text{if}\;\; \text{OW},
\end{array} \right.
\end{eqnarray*}
where \[\mathcal{N}_{cts}=\int_{0}^{2\mu}\frac{1}{\sqrt{2\pi\sigma^2}}e^{-\frac{({\scriptstyle\mathcal{S}}-\mu)^2}{2\sigma^2}}d{\scriptstyle\mathcal{S}}.\]
We will use this formulation in what follows.

\subsection*{Mutual information calculations}
In this section we apply information-theoretic measures introduced by Shannon \cite{BLTJ:BLTJ1338} to quantify the ability of synapses to transmit information. One important concept in information theory is \textbf{mutual information}. It measures the expected reduction in uncertainty about $x$ that results from learning $y$, or vice versa. This quantity can be formulated
\[I\left(X;Y\right) = H\left(X\right)+H\left(Y\right)-H\left(X,Y\right),\]
where \textbf{entropy} was defined earlier and
\begin{eqnarray}
H\left(X,Y\right)& = -\sum_{y\in\mathcal{Y}}\sum_{x\in\mathcal{X}}p\left(X=x,Y=y\right)\log_{2}p\left(X=x,Y=y\right)\label{jointent}
\end{eqnarray}
where the \textbf{joint entropy} of two random variables $X$ and $Y$ quantifies the uncertainty of their joint distribution.
Using Eqs. (\ref{margent}) and (\ref{jointent}), the mutual information can be rewritten
\begin{eqnarray}
I(X;Y)=\sum_{x\in\mathcal{X}}\sum_{y\in\mathcal{Y}}p(x,y)\log_{2}\frac{p(x|y)}{p(x)}.
\end{eqnarray}
The mutual information is symmetric in the variables $X$ and $Y$, $I(X;Y)=I(Y;X)$, and is zero if the random variables are independent or if the relation between them is deterministic (nothing to be learned in either case).
 In order to compute this from data one is faced with the basic statistical problem of estimating the entropies: selecting the correct bin width in the histograms of the random variables. Here, we use \textit{Freedman-Diaconis} rule \cite{Freedman1981} for finding the optimal number of bins. According to this rule, the optimal number of bins can be calculated base on interquartile range $(Iqr = Q_3-Q_1)$ and the number of data points $n$. It is defined as
\[{nbins} = \frac{{max}(x)-{min}(x)}{2*Iqr*n^{-1/3}}.\]
One of the advantages of  \textit{Freedman-Diaconis Rule} is that the variability in data is taken into account using $Iqr$, which is resistant to outliers in the data.

We stimulated synapses with input Poisson spike trains with different mean frequencies and compute the mutual information between postsynaptic response and the preceding interspike interval. The mutual information then is obtained by
\begin{eqnarray*}
I\left(\mathcal{S};T\right) = H\left(\mathcal{S}\right)+H\left(T\right)-H\left(\mathcal{S},T\right).
\end{eqnarray*}

\begin{figure}[H]
   \includegraphics[scale=0.65]{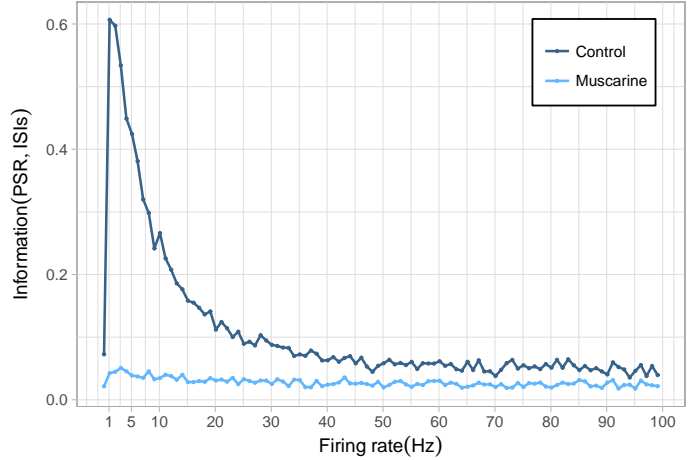}
  \caption{Mutual Information between postsynaptic response and preceding inter-spike intervals for the control and muscarine parameter sets.}\label{MIPSRISI}
      \end{figure}

Figure \ref{MIPSRISI} shows the result of this calculation for the control and muscarine cases. In both we observe a rapid decline in mutual information at frequencies above 7 Hz. The peak mutual information occurs between 0.1 to 2 Hz,  all of which demonstrate frequency dependence of temporal information encoding.  Large inter-spike intervals (low frequency) allow enough time for recovery of all release sites, leading to a tight distribution about $Pr=P_{max}$, and very little communication of information.   Very short inter-spike intervals give no time for recovery and $Pr$ is confined to a small region near 0, also leading to low information transmission. Between these extremes, variable inter-spike intervals have a significant effect on the amplitude of $Pr$.  The main difference between the muscarine and control cases is the absolute value of the mutual information, which is significantly lower in muscarine case. This can be understood by considering that the range of $Pr$ values in the muscarine case is much smaller, so variations in $Pr$ are less distinguishable, and contribution less to the mutual information.
This coincides with results in \cite{Fuhrmann140}, where depressing synapses are found to have a peak mutual information at very low frequencies.  However, here we also see this in the distributions themselves, where the amount of variability is represented by the overall width or entropy of the distribution.  It is not surprising that the mutual information will reflect this, especially considering that these entropies are the foundation of the calculation.

\section*{Information ``Stored" in the Postsynaptic Response}
We showed that a distribution of  postsynaptic responses (PSR) carries information about the distribution of the ISIs (interspike intervals). However, the exact sequence of ISIs determines a given postsynaptic response.  The length of the sequence that effects a response will depend on the parameters of the model, in that the calcium can accumulate in certain situations. We call this a kind of ``memory" for an exact sequence of ISIs. We measure this by computing the mutual information between a prior sequence, and the PSR.  This can be done in coarse grained way by adding up a sum of $N$ previous ISIs and using that as a random variable, which we consider first.  This, however, ignores the subtleties of the exact sequence; for instance, a long ISI followed by a short ISI will give a PSR that is smaller than the reverse, even if the sum of the two is the same.  A more complete approach is to compute the mutual information between a PSR and an $N$-tuple of preceding spikes, in order.  The former method has been used in previous work so we begin with that approach to demonstrate some of these aforementioned subtleties.

\subsection*{MI between PSR and sum of preceding ISIs}
 In \cite{Fuhrmann140}, they compute the mutual information between the postsynaptic responses a sum of the preceding ISIs, increasing the number in the sum to go further back in time.
Assuming that first spike occur at $t_0=0$ ms, this can be formulated as follows.
Let $ t_{1}=T_1, t_{2}=T_{1}+T_{2},\cdots,t_{k}=T_{1}+\cdots+T_{k}$ be a vector of the sums of the preceding ISIs. It can be simplified
\[t_{k}=\sum_{i=1}^{k}T_{i}\;\;\;\text{for}\;\; k=1,\cdots, M,\]
where $M>0$ is a natural number. The mutual information between the postsynaptic response and the  sum of preceding spike ISIs is then
\[I(\mathcal{S};t_{k})=H(\mathcal{S})-H(\mathcal{S} \mid t_{k})\;\;\;\text{for}\;\; k=1,\cdots, M\]

\begin{figure}[H]
   \includegraphics[scale=0.65]{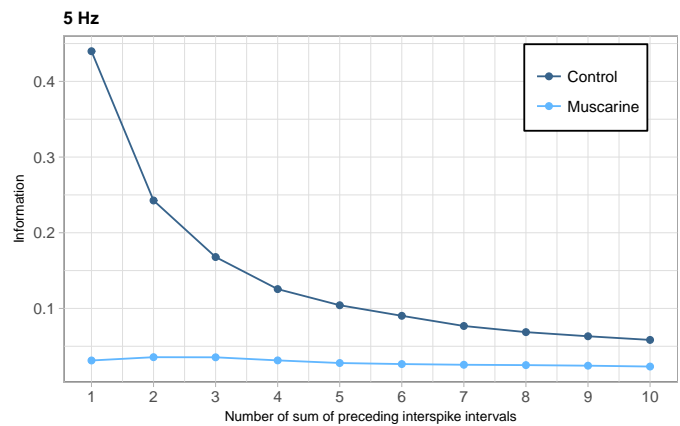}
   \caption{Information between postsynaptic response and number of sum of preceding interspike intervals in the control and muscarine cases for firing rate $5$ Hz.}\label{infosum5}
  \end{figure}

\begin{figure}[H]
   \includegraphics[scale=0.65]{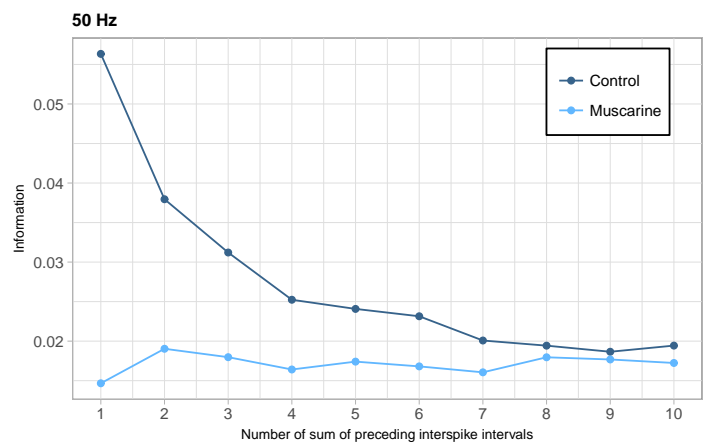}
  \caption{
      Information between postsynaptic response and number of sum of preceding interspike intervals in control and muscarine cases for firing rate $50$ Hz.}\label{infosum50}
      \end{figure}

Figures \ref{infosum5} and \ref{infosum50} illustrate the results for both the control and muscarine cases, at $5$ and $50$ Hz, respectively. The information contained in the sum is significantly lower for muscarine compared to the control case, which is not surprising, given the results from the preceding section. For both firing rates, $5$ and $50$ Hz, in the control case the MI decreases as more ISIs are added to the sum.  The further back in time the sum goes, the less the ISIs in the sum are directly involved in determining the PSR.  The MI curve becomes almost flat past 5 preceding ISIs for $50$ Hz, indicating an effect that can be measured only that far back.  In the control $5$  Hz case the decay begins to level off at $N=5$ as well, but continues to decline for $N>5$.  This shows a frequency dependence of the measure that makes sense mechanistically.  The muscarine MI is much less dependent on the cumulative history of spikes, showing very little change as N is increased.  This also makes sense mechanistically, because in the muscarine case the response range is very narrow, and cannot carry much information forward from one preceding ISI, let alone any ISIs preceding that.

\subsection*{Mutual information (MI) between postsynaptic response $\mathcal{S}$ and $n$-tuple  inter-spike intervals $T_1,T_2,\cdots,T_n$ }
Our second method is much more computationally intense, but preserves the exact structure of the sequence of preceding ISIs.

Consider a single-input and single-output channel with inter-spike input $T$ and postsynaptic response output $\mathcal{S}$. The mutual information between $T$ and $\mathcal{S}$ in this notation is defined
\begin{eqnarray*}
I(T;\mathcal{S})=H(T)+H(\mathcal{S})-H(T,\mathcal{S}).
\end{eqnarray*}
Consider now a channel with two interspike interval inputs $T_{1}$ and $T_{2}$ and a single output $\mathcal{S}$.  The mutual information between the inputs and the output of two-way channel is commonly known as the $2$-tuple information and can be defined as the amount of reduction of uncertainty in $\mathcal{S}$ knowing $(T_1,T_2)$.  The $2$-tuple  mutual information is written
\begin{eqnarray*}
I(<T_1,T_2>;\mathcal{S})=H(T_1,T_2)+H(\mathcal{S})-H(T_1,T_2,\mathcal{S}).
\end{eqnarray*}
 Following McGill \cite{McGill1954} we can extend the definition for mutual information to include more than two sources $T_1, T_2,\cdots,T_{n}$ that transmit to
  $\mathcal{S}$, arriving at
  \begin{eqnarray}
I(<T_1,T_2,\cdots,T_{n}>;\mathcal{S})= H(\mathcal{S})+H(T_1,T_2,\cdots,T_{n})-H(T_1,T_2,\cdots,T_{n},\mathcal{S}).
  \end{eqnarray}

The histogram construction of probability density functions in higher dimensions suffers from the ``curse of dimensionality" \cite{Kwak2002}, for obvious reasons. Kozachenko and Leonenko \cite{Kozachenko1987} instead proposed an improved nonparametric estimator for entropy: the KL estimator based on $k$-nearest neighbor distances.  In \cite{PhysRevE.69.066138} Kraskov et al. reported on a modified KL estimator to compute mutual information with improved performance and applicability in higher dimensional spaces (the result is referred to as the KSG algorithm). Compared to other methods for multivariate data sets, estimators obtained by the KSG algorithm have minimal bias when applied to data sets of limited size, as in common in real world problems.

The KL estimator demonstrates that it is not necessary to estimate the probability density function in order to compute information theoretic functionals. Instead, the estimator is based on the metric properties of nearest neighbor balls in both the joint and marginal spaces. The general form of the KL estimator for differential entropy is written as
\begin{eqnarray*}
\hat{H}(X) = \psi(N)-\psi(k)+\log c_{d}+\frac{d}{N}\sum_{i=1}^{N}\log\epsilon(i),
\end{eqnarray*}

where $\psi:\mathbb{N}\rightarrow\mathbb{R}$ denotes the digamma function, $\epsilon(i)$ is twice the distance from point $x_i$ to its $k$-th neighbor, $d$ is the dimension of $x$ and $c_d$ is the volume of the $d$-dimensional unit ball. Similarly, KL estimator for the joint entropy is written as
 \begin{eqnarray}
 \hat{H}(X,Y) = -\psi(k)+\psi(N)+\log c_{d_X} c_{d_Y}+\frac{d_X+d_Y}{N}\sum_{i=1}^{N}\log\epsilon(i).
 \end{eqnarray}
 Therefore, for any fixed $k$,  the mutual information can be computed by subtracting the joint entropy estimate from $H(X)$ and $H(Y)$.  However, this introduces biases due to the different distance scales in different dimensions.  The KSG algorithm corrects this by instituting a random choice of the nearest neighbor parameter $k$.  For a more detailed derivation see \cite{PhysRevE.69.066138}.

We now employ it to estimate mutual information between the probability of release ($Pr$) as single-output and preceding ISIs as a multivariate input. Note that we are able to use the deterministic $Pr$ distributions in these calculations because $Pr$  is not directly dependent upon the n-tuple of the preceding ISIs.

\begin{figure}[H]
  \begin{center}
   \includegraphics[scale=0.65]{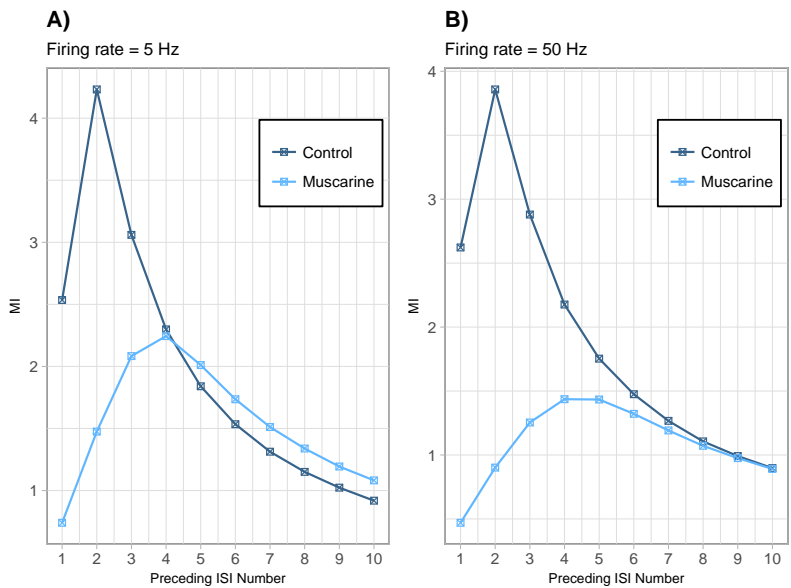}
  \caption{
     N-tuple information between inter-spike intervals and $Pr$ for control and muscarine cases for two frequencies: $5$ and $50$ Hz.}\label{tupleboth}
     \end{center}
      \end{figure}

Figure \ref{tupleboth} shows the increase in mutual information (or reduction in uncertainty) between the $Pr$ and an n-tuple ISI, with increasing $n$. Mean rates in the gamma and theta ranges, $5$ and $50$ Hz, respectively, are plotted for the control and muscarine cases. At 5 Hz in the control case the mutual information increases from 1 to 2 preceding ISIs, after which it decreases, apparently exponentially.  Because the MI is still greater than for 1 ISI for 3 preceding ISIs, we could say that this synapse ``remembers" up to 3 preceding ISIs.  For the muscarine case the increase happens over the first 4 preceding ISIs, meaning the uncertainty is reduced by adding in previous ISIs, and in this parameter regime the synapse ``remembers" somewhat further back in the ISI train than in the control case.   The mutual information is even greater in muscarine than control for more than 4 ISIs at 50 Hz. This is not at all obvious from the other results we have presented in this paper, and could not be uncovered without these statistics on sequences of ISIs. At 50 Hz the muscarine MI is always smaller than the control MI, but we see almost the same history dependence for each synapse independent of frequency.

\section*{Discussion \& Conclusion}

When presented with a Poisson spike train input, our depressing synapse model acts as a nonlinear filter of the exponential distribution of inter spike intervals.  For high and low input frequencies, the result is as expected, the output $Pr$ distribution is peaked at values near zero and $P_{max}$ respectively, with very small variance (see Figure \ref{FreqDist2}).  In between, for frequencies near theta, the distribution is spread across the entire interval between zero and $P_{max}$.  This creates a peak in the entropy of the distribution near this value, which demonstrates a wide dynamic range in the response of the synapse (Figure \ref{PrEntropy}).  The range in the control case is larger, necessarily, because $P_{max}$ is larger.

Over the gamma frequency range we see that the peak and mean of the distribution remains more or less constant, indicating a stable response size when presented with a Poisson spike train. This would create a stable connection and is the advantage of having a peaked distribution with low entropy.  The addition of muscarine reduces the strength of this connection.

These are the results of a numerical investigate of the properties of the synapse.  Creating closed form expressions for these distributions in not possible, the map is too complex.  However, the stochastic recurrence relation for the calcium concentration alone is simple enough to be analyzed and we discovered that it follows a gamma distribution with shape parameter $\lambda \tau_{ca} +1$.  For the $Pr$ distribution we had to rely on an approximation motivated by numerical results.  We found the mean of the distribution followed the frequency in the same way that the fixed point of the map did.  The collapse to the fixed point is very rapid, one or two iterations at most, which justifies our assumption that the $Pr$ values are determined by the fixed point value associated with the instantaneous rate of the preceding inter spike interval, e.g. $\lambda = T$.  Then the formula for the fixed point as a function of frequency could be used to generate a distribution, using the exponential distribution of the $T$'s.  The results confirm the validity of this approximation, even in ranges outside the other approximations necessary to arrive at a closed form.  Most importantly, the ``sloshing" of the distribution between zero and $P_{max}$ as the frequency is decreased through the physiological range is captured by this form.

For a given train of inter spike intervals the map for $Pr = P R$ is completely deterministic, it captures the mean of the response behavior.  Therefore the mutual information between the $Pr$ distribution and the ISI distribution should be zero.  In order to introduce stochasticity to the response we modeled the noise in the release of vesicles of neurotransmitter as a Binomial process, and the noise postsynaptic response to the vesicles with a truncated Normal distribution.  With a distribution of the resulting PSR values, we can calculate mutual information.  The mutual information as a function of firing rate was found to have a peak around 3 Hz for both the muscarine and the control cases, though  the over all mutual information was much lower for muscarine, in part because of the lower entropy of  the $Pr$ distribution in this case (Figure \ref{MIPSRISI}).  The peak in the mutual information in the control case is 6 times that of the baseline value for large firing rates.  For the muscarine case it was only about 1.25 times larger.  Therefore the synapses with muscarine added as a neuromodulator are much less sensitive to changes in frequency over all.  This can be viewed as a good thing with regard to stability of the response, but definitely allows for a much less dynamic filter than in the control case.

We then took these calculations a step further by attempting to determine how far back in a spike train the synapses ``remembers", in that information from previous ISIs are carried forward into a certain measurement of the $Pr$.  This is a non-trivial problem, with technical challenges in the calculations.  We performed a mutual information calculation with a distribution of the sum of previous ISIs, including successively more times.  This sum collapses much of the information in the sequence, measuring only if it was overall a long time or a short time, compared to other samples.  Some results are obtainable, however, and we saw an over all decline in mutual information as more ISIs were added to the sum, at least in the control case (fig.s \ref{infosum5}, \ref{infosum50}).  In the muscarine case the mutual information is seemingly insensitive to adding in more ISIs, being relatively flat.  The absolute value of the mutual information was much less in the both cases for 50 Hz than for 5 Hz.  Again, the larger entropy of the $Pr$ distribution at 5 Hz creates the potential for larger mutual information.  We did see that the MI leveled out more slowly at 5 Hz than 50 Hz, too, for the control case, indicating a longer ``memory" at 5 Hz than at 50.

In an attempt to keep all the structure in a preceding inter spike interval sequence we used a multivariate version of the mutual information calculation.  Because of the high dimensionality of the data this is more challenging.  A k-nearest neighbor approach to estimating entropy is appropriate in this case and we used the KSG algorithm to do the calculation.  This measures the mutual information between the response and the sequence of ISIs as the number in the sequence is increased.  If it increases with more ISIs in the sequence we can assume that the reduction of uncertainty is greater as longer histories are considered. It is seen to increase initially and then decrease with history length as the memory effect is washed out (Figure \ref{tupleboth}).  Somewhat non-intuitively the muscarine case increases over longer histories than in control, though the over all MI is again smaller.  Adding up to 4 ISIs to the sequence improves the prediction of the $Pr$ vs. 2-3 ISIs in control.  This result is more or less the same at gamma and theta frequencies.  Using the sum of the ISIs in the mutual information calculation hides this dependence in the muscarine case, as anticipated.

Through this analysis we've gained insight into the information processing characteristics of PV BCs.  Within a physiological context, PV BCs receive synaptic input from intrinsic and extrinsic sources, effectively tuning them to fire specifically at theta frequency \cite{Varga2012}.  It is clear from our quantitative description that constraining the firing of PV BCs at theta frequency optimizes the information content of PV BC to pyramidal cell synaptic transmission.  Moreover, PV BCs in vivo often burst more than one spike per theta cycle \cite{Varga2012}.  This short ``burst" of more than one action potential in vivo serves to further optimize the information content, providing a stronger ``memory" of PV BC activity.

Cholinergic neuromodulation of PV BCs occurs both pre- and postsynaptically (\cite{Lawrence2408}, \cite{Yi}) and is associated with the generation of population-level gamma rhythms.  By reducing $Pr$ and protecting the synapse from vesicular depletion, we now show that presynaptic neuromodulation reduces entropy and mutual information.  While this effect may be reduce the information content of the synapse, it will promote stability and regularity of the synaptic response that might be advantageous in the generation of population-level gamma rhythms.  Therefore, we hypothesize two modes for PV BC synapses.  In the absence of cholinergic neuromodulation, PV BCs are optimally tuned to transfer information at theta frequency, which might be advantageous in encoding the onset of salient stimuli \cite{Pouille2004}.  In the presence of cholinergic neuromodulation, when PV BCs may become depolarized \cite{Yi} and are more likely to fire at gamma frequency, the information processing capability is reduced, gaining synapse stability, which would promote population-level network synchronization.  Future studies that employ ``in vivo" spike trains could corroborate these hypotheses.

We are continuing our quest to quantify the information processing properties of synapses using more novel techniques than mutual information.  Computational mechanics (\cite{Crutchfield94thecalculi}, \cite{Shalizi2001})  gives us a theory and a basis for understanding time series from a process as a hidden Markov model with multiple states.  How the model changes as input frequency is varied, or neuromodulation is applied, would give us an even better categorization of the synapse processes.  Furthermore, these measures can be applied to output processes alone, without any information about the input stimuli, allowing a much more robust classification of signals measured from synapses in vivo or in vitro.

Understanding the interaction of the synaptic dynamics with voltage oscillations in the hippocampus is the next step to connecting the dots between synaptic plasticity and it effect on the mechanisms involved in learning and memory. We have preliminary results in a simple deterministic model and will be extending these to stochastic models of micro-circuit rhythms.  We believe that a careful piecing together of the model and behavior will guide our intuition much more than a large scale numerical approach, and will more easily interface with electrophysiological experiments on actual neurons in the hippocampus.

\section*{Appendices}
\subsection*{Appendix A}

Let $\theta_1,\theta_2,\cdots$ be iid $\mathbb{R}^2$-valued ($d\geq 1$) random variables with generic copy $\theta$ and independent of $X_{0}$. Suppose that $X_{n}=\Psi\left(X_{n-1},\theta_n\right)$ for all $n\geq 1$ and a continuous function $\Psi : \mathbb{R}^{d+1}\rightarrow\mathbb{R}$. If $X_n$ converges in distribution to $X_{\infty}$, then
\[\Psi\left(X_{n-1}, \theta_n\right)\stackrel{d}{\rightarrow}\Psi\left(X_{\infty},\theta\right)\;\;\;\text{and}\;\;\; X_{\infty}\stackrel{d}{=}\Psi\left(X_{\infty},\theta\right) \]

\subsection*{Appendix B}

Suppose independent, identically distributed increases in the amount of calcium $\{\Delta_{n}\}_{n\geq1}$ occur at times $\{t_n\}_{n\geq1}$, and we are interested in the distribution of the amount of calcium accumulated. In one dimension, the random recurrence equation for the calcium concentration is given by
\begin{eqnarray}
C_{n} = A_{n}C_{n-1}+\Delta_{n},\hspace{5mm}n\geq1,\label{Cn}
\end{eqnarray}
where $A_{n}= e^{-T_{n}/\tau_{ca}}$, and the waiting times $T_1=t_1$, $T_n = t_{n}-t_{n-1}$, $n\geq2$, are i.i.d., making the $\{t_n\}$ a renewal process. Moreover, $\left(A_{n},\Delta_{n}\right)$ are assumed to be i.i.d. vectors. $C_{0}$ is the base calcium concentration which is assumed to be zero because near the calcium sensors the concentration is extremely low in the absence of a spike.

Iterating  (\ref{Cn}) leads to
\begin{align*}
C_{n}&=A_{n}C_{n-1}+\Delta_{n},\\
&=A_{n}A_{n-1}C_{n-2}+A_{n}\Delta_{n-1}+\Delta_{n},\\
&= A_{n}A_{n-1}A_{n-2}C_{n-3}+A_{n}A_{n-1}\Delta_{n-2}+A_{n}\Delta_{n-1}+\Delta_{n},\\
&\vdots\\
&=A_{n}A_{n-1}.\cdots.A_{1}C_{0}+\sum_{k=1}^{n}A_{n}.\cdots.A_{k+1}\Delta_{k}
\end{align*}
for each $n\geq1$.
Using the independence assumptions and replacing $\left(A_{k},\Delta_{k}\right)_{1\leq k\leq n}$ with the copy $\left(A_{n+1-k},\Delta_{n+1-k}\right)_{1\leq k\leq n}$ we observe that
\begin{eqnarray*}
C_{n}\stackrel{d}{=} A_{n}A_{n-1}.\cdots.A_{1}C_{0}+\sum_{k=1}^{n}A_{1}A_{2}.\cdots.A_{k-1}\Delta_{k}
\end{eqnarray*}
where $\stackrel{d}{=}$ denotes equality in distribution.

A fundamental theoretical result shown in \cite{kesten1973} asserts that if
\begin{eqnarray*}
E\left(\ln|A|\right)<0\hspace{5mm}\text{and}\hspace{5mm}E\left(\ln|\Delta|\right)<\infty
\end{eqnarray*}
then the series
\begin{eqnarray*}
C=\sum_{k=1}^{\infty}A_{1}A_{2}.\cdots.A_{k-1}\Delta_{k},
\end{eqnarray*}
will converge w.p.1 and the distribution of $C_n$ converges to that of $C$, independently of $C_{0}$. Also, by the Continuous Mapping Theorem (see Appendix A), we infer from (\ref{Cn}) that if $C_{n}\stackrel{d}{\rightarrow}C$, then $C$ satisfies the distributional identity
\begin{eqnarray*}
C\stackrel{d}{=}AC+\Delta,\hspace{5mm}C\;\text{and}\;\left(A,\Delta\right)\hspace{2mm}\text{independent}
\end{eqnarray*}
Following \cite{vervat} let $X:=AC$, then we can write
\begin{eqnarray}
X\stackrel{d}{=}A\left(X+\Delta\right), \label{X}
\end{eqnarray}
and hence
\begin{eqnarray*}
C\stackrel{d}{=}X+\Delta.
\end{eqnarray*}
Iterating (\ref{X})  results in
\begin{eqnarray*}
X\stackrel{d}{=}\sum_{n=1}^{\infty}A_{1}A_{2}.\cdots.A_{n}\Delta_{n}
\end{eqnarray*}
where $A_{n} = e^{-T_{n}/\tau_{ca}}$.

Note that if $T_{n}$ is exponentially distributed with rate parameter $\lambda$, then random variable $A_n$ has Beta distribution with shape parameters ($\lambda\tau_{ca}$,$1$) and is denoted by $A_{n}\sim \textit{Beta}\left(\lambda\tau_{ca},1\right)$.
We then apply a beta-gamma algebra identities (\cite{DUFRESNE1998285}) which state that
\begin{eqnarray*}
Beta\left(a,b\right)\odot Gamma\left(a+b,1\right)=Gamma\left(a,1\right).
\end{eqnarray*}
Alternatively,
\begin{eqnarray}
Beta\left(a,b\right)\odot\left( Gamma\left(a,1\right)+Gamma\left(b,1\right)\right)\stackrel{d}{=}Gamma\left(a,1\right).\label{Beta}
\end{eqnarray}
Applying (\ref{Beta}) in (\ref{X}), and considering that $A_{n}\sim \textit{Beta}\left(\lambda\tau_{ca},1\right)$, then
\begin{align*}
X&\sim Gamma\left(\lambda\tau_{ca},1\right),\\
\Delta&\sim Gamma\left(1,1\right).
\end{align*}
note that $C\stackrel{d}{=}X+\Delta$ implies
\begin{eqnarray*}
C\stackrel{d}{=}Gamma(\lambda\tau_{ca},1)+Gamma(1,1).
\end{eqnarray*}
Finally
\begin{eqnarray*}
C\sim Gamma(\lambda\tau_{ca}+1,1).
\end{eqnarray*}
Note that a random variable $X$ that is gamma-distributed with shape $\alpha$  and scale $\theta$ parameters is denoted by $Gamma\left(\alpha,\theta\right)$ and the corresponding probability density function  is
\begin{eqnarray*}
f\left(x;\alpha,\theta\right) = \frac{1}{\Gamma(\alpha)\theta^{\alpha}}x^{\alpha-1}e^{-\frac{x}{\theta}},\;\;\;\text{for}\;\;x>0\;\text{and}\;\alpha, \theta>0
\end{eqnarray*}
Also, a random variable $X$ that is beta-distributed with shape parameters $\alpha$  and  $\beta$  is denoted by $Beta\left(\alpha,\beta\right)$ and the probability density function of the beta distribution is
\begin{eqnarray*}
f\left(x;\alpha,\beta\right) = \frac{1}{B(\alpha,\beta)}x^{\alpha-1}(1-x)^{\beta-1},\;\;\;\text{for}\;\;0<x< 1\;\text{and}\;\alpha, \beta>0
\end{eqnarray*}
where $B(\alpha,\beta)=\frac{\Gamma(\alpha)\Gamma(\beta)}{\Gamma(\alpha+\beta)}$.
\subsection*{Appendix C}
 Computing an approximate distribution for $Pr$
For the ease of notation, let $X=\overline{R}$ be a random variable defined as follows
\begin{eqnarray*}
X= \frac{1-e^{-k_{min}T}}{1-(1-P_{max})e^{-k_{min}T}}.
\end{eqnarray*}
Let the random variable $T$ have the exponential distribution with probability density function
\begin{eqnarray*}
f_{T}(t) = \lambda e^{-\lambda t}\hspace{5mm}t>0
\end{eqnarray*}
We can compute an analytical expression for probability density function (PDF) of fixed point $\overline{R}$ using the distribution for $T$.

The transformation $X = g(T) =  \frac{1-e^{-k_{min}T}}{1-(1-P_{max})e^{-k_{min}T}}$ is a 1-1 transformation from $\mathcal{T} = \{t|\; t > 0\}$ to $\mathcal{X} = \{x|\;0 <x<1\}$ with inverse $T= g^{-1}(X) = \frac{1}{k_{min}}\log\left(\frac{1-cx}{1-x}\right)$ and Jacobian
 \[\frac{dT}{dX}=\frac{1-c}{k_{min}(1-x)(1-cx)}\]
 By the rule for functions of random variables, the probability density function of $X$ is
 \begin{align*}
 f_{X}(x) &= f_{T}\left(g^{-1}(x)\right)\left|\frac{dt}{dx}\right|\\
 &= \frac{\lambda\left(1-c\right)}{k_{min}}\left(1-x\right)^{-\left(1-\lambda/k_{min}\right)}\left(1-cx\right)^{-\left(1+\lambda/k_{min}\right)}
 \end{align*}

 \section*{Abbreviations}
  BC-basket cell\\ \noindent
  CA1- Cornu Ammonis, earlier name for hippocampus\\\noindent
 FD-facilitation and depression\\\noindent
 IPSC-inhibitory postsynaptic current\\\noindent
 ISI-interspike interval\\
 KL-Kozachenko and Leonenko\\
 KSG-Kraskov, St\"{o}gbauer, and Grassberger\\
 mAChR-muscarinic acetylcholine receptors\\
 MCMC-Monte Carlo Markov Chain\\
 MI-Mutual Information\\
 NT-neurotransmitter\\
 PSR-postsynaptic response\\
 PV-parvalbumin-positive\\

\section*{Authors' contributions}
EBM did all the statistical analysis in the study. JL carried out all experiments and preliminary data analysis.  ES conceived of the study, developed the design and analyzed the results.
    All authors read and approved the final manuscript.
     \section*{Authors' email addresses}
 \noindent
 Elham Bayat-Mokhtari:  Elham.bayatmokhtari@umontana.edu\\
 \noindent
 J.Josh Lawrence: john.lawrence@ttuhsc.edu\\
 \noindent
 Emily Stone: stone@mso.umt.edu\\
    \section*{Availability of data and materials}Please contact author for data requests.
    \section*{Competing interests}
  The authors declare that they have no competing interests.

    \section*{Consent for publication}Not applicable
    \section*{Ethics approval and consent to participate}Not applicable
    \section*{Funding}
    Electrophysiology experiments were performed in the laboratory of Chris McBain with intramural support from National Institute of Child Health and Human Development. Later work was supported by National Center for Research Resources Grant P20-RR-015583, National Institutes of Health Grant R01069689-01A1, and start-up support from the University of Montana Office of the Vice President for Research (to J. J. Lawrence).


\setcounter{figure}{0}


\section*{Tables}
\begin{table}[h!]
\caption{Parameters in the map}{\label{tble:1}}
      \begin{tabular}{rl}
        \hline
Parameter   &  Description \\ \hline
 $\Delta$ & Increase in the amount of calcium relative to  control conditions\\
$P_{max}$ & Maximum probability of release \\
 $K$   & Half calcium concentration value for probability of release function\\
$k_{min}$ & Minimum rate of recovery of synapses\\
$k_{max}$ & Maximum rate of recovery of synapses\\
 $K_{r}$ & Half calcium concentration value for rate of recovery function\\
 $\tau_{Ca}$ & Decay constant for calcium\\ \hline
      \end{tabular}
\end{table}

\begin{table}[h!]
\caption{Parameter values}\label{tble:2}
      \begin{tabular}{rl}
        \hline
Parameter   &  Fitted value \\ \hline
 $K$ & $0.2$\\
$k_{min}$ & $0.0017$ 1/msec  \\
 $k_{max}$   & $0.0517$ 1/msec\\
$K_r$ & $0.1$\\
$\tau_{Ca}$ & $1.5$ msec\\  \hline
      \end{tabular}
\end{table}



\section*{Acknowledgments}
  We acknowledge David Patterson for his helpful comments on some of the statistical techniques.

\nolinenumbers

\bibliography{library}

\bibliographystyle{abbrv}

\end{document}